\documentclass[sigconf,screen]{acmart}

\usepackage{breakurl}
\usepackage{enumitem}

\AtBeginDocument{%
  \providecommand\BibTeX{{%
    \normalfont B\kern-0.5em{\scshape i\kern-0.25em b}\kern-0.8em\TeX}}}

\copyrightyear{2023} 
\acmYear{2023} 
\setcopyright{rightsretained} 
\acmConference[FAccT '23]{2023 ACM Conference on Fairness, Accountability, and Transparency}{June 12--15, 2023}{Chicago, IL, USA}
\acmBooktitle{2023 ACM Conference on Fairness, Accountability, and Transparency (FAccT '23), June 12--15, 2023, Chicago, IL, USA}
\acmDOI{10.1145/3593013.3594044}
\acmISBN{979-8-4007-0192-4/23/06}

\begin{document}

\title{Algorithmic Unfairness through the Lens of EU Non-Discrimination Law}
\subtitle{Or Why the Law is not a Decision Tree}

\author{Hilde Weerts}
\authornote{Both authors contributed equally to this research.}
\affiliation{%
  \institution{Eindhoven University of Technology}
  \country{The Netherlands}}
\email{h.j.p.weerts@tue.nl}

\author{Rapha\"{e}le Xenidis}
\authornotemark[1]
\affiliation{%
  \institution{Sciences Po Law School}
  \country{France}}
\email{raphaele.xenidis@sciencespo.fr}

\author{Fabien Tarissan}
\affiliation{%
  \institution{CNRS \& ENS Paris-Saclay}
  \country{France}}
\email{fabien.tarissan@ens-paris-saclay.fr}

\author{Henrik Palmer Olsen}
\affiliation{%
  \institution{University of Copenhagen}
  \country{Denmark}}
\email{henrik@jur.ku.dk}

\author{Mykola Pechenizkiy}
\affiliation{%
  \institution{Eindhoven University of Technology}
  \country{The Netherlands}}
\email{m.pechenizkiy@tue.nl}


\begin{abstract}
Concerns regarding unfairness and discrimination in the context of artificial intelligence (AI) systems have recently received increased attention from both legal and computer science scholars. Yet, the degree of overlap between notions of algorithmic bias and fairness on the one hand, and legal notions of discrimination and equality on the other, is often unclear, leading to misunderstandings between computer science and law.  
What types of bias and unfairness does the law address when it prohibits discrimination? What role can fairness metrics play in establishing legal compliance? In this paper, we aim to illustrate to what extent European Union (EU) non-discrimination law coincides with notions of algorithmic fairness proposed in computer science literature and where they differ.
The contributions of this paper are as follows. First, we analyse seminal examples of algorithmic unfairness through the lens of EU non-discrimination law, drawing parallels with EU case law. 
Second, we set out the normative underpinnings of fairness metrics and technical interventions and compare these to the legal reasoning of the Court of Justice of the EU. Specifically, we show how normative assumptions often remain implicit in both disciplinary approaches and explain the ensuing limitations of current AI practice and non-discrimination law. 
We conclude with implications for AI practitioners and regulators.
\end{abstract}



\settopmatter{printfolios=true}
\maketitle

\section{Introduction}
Concerns regarding algorithmic unfairness and discrimination are receiving increased attention from both legal and computer science scholars. Yet, the degree of overlap between computer sciences notions of bias and fairness and legal notions of discrimination and equality is often unclear. On the one hand, computer scientists have put forward various metrics and technical interventions to measure and mitigate unfairness of artificial intelligence (AI) systems. However, an AI practitioner hoping for an explicit answer to the question: "what should be the value of my fairness metric for my system to be compliant with the law?" is likely to be disappointed, as most of the time the answer will amount to a variation of "it depends". On the other hand, challenges of algorithmic unfairness are not always properly understood by legal scholars. As a result, legal experts and regulators struggle with figuring out how discrimination law can properly address algorithmic bias and unfairness. Moreover, there exists a tendency in the legal community to overestimate the effectiveness and applicability of technical interventions \citep{balayn2021beyond}.

This raises several important questions. What types of bias and unfairness does the law address when it prohibits discrimination? What role can fairness metrics play in establishing legal compliance -- if any? This paper aims to respond to computer scientists’ uncertainties about what is legal when it comes to discrimination, and to lawyers’ questions regarding the challenges and technical possibilities to realise equality rights and non-discrimination law obligations. To this end, we show to what extent non-discrimination law coincides with notions of algorithmic fairness proposed in computer science literature and where they differ. 

Existing work in this direction has primarily targeted a legal audience \citep[e.g.][]{binns2020algorithmic,wachter2020bias,wachter2021fairness}. Most notably, \citet{wachter2021fairness} set out how the contextual nature of EU non-discrimination law makes it impossible to automate non-discrimination in the context of AI systems and propose a fairness metric that aligns with the Court’s "gold standard". Additionally, several works focus on US anti-discrimination law~\citep[e.g.][]{hellman2020measuring,kim2022race,kumar2022equalizing}. For example, \citet{hellman2020measuring} considers the compatibility of several fairness metrics under US anti-discrimination law and touches upon the legitimacy of particular types of technical interventions.

In this paper, we consider European Union (EU) non-discrimination law and target a broader audience, bridging two distinct disciplines. The contributions of this paper are as follows. Following a brief introduction to EU Discrimination law, we analyse seminal examples of algorithmic unfairness through the lens of EU non-discrimination law, drawing parallels with EU case law. 
Second, we set out the normative underpinnings of fairness metrics and technical interventions and compare these to the legal reasoning of the court. Specifically, we show how normative assumptions often remain implicit in both disciplinary approaches and explain the ensuing limitations of current AI practice and non-discrimination law. 

The remainder of the paper is structured as follows. 
Section~\ref{sec:law} provides the necessary background on EU non-discrimination law. Section~\ref{sec:comparison} presents our analysis of seminal examples from the algorithmic fairness literature through the lens of EU non-discrimination law. Building on these findings, Section~\ref{sec:normative} explores the normative underpinnings of fairness metrics, fairness-aware machine learning algorithms, and the legal reasoning of the Court of Justice of the EU. In Section~\ref{sec:discussion}, we discuss the implications of our findings for AI practitioners and regulators and Section~\ref{sec:conclusion} concludes the paper.

\section{Discrimination under EU law}
\label{sec:law}
Following \citet{lippert2006badness}, discrimination can generally be characterised by the morally objectionable practice of subjecting a person (or group of persons) to a treatment in some social dimension that, for no good reason, is disadvantageous compared to the treatment awarded to other persons who are in a similar situation, but who belong to another socially salient group.\footnote{\citet{lippert2006badness} considers a group to be socially salient "if perceived membership of it is important to the structure of social interactions across a wide range of social contexts".} Central to this definition is the comparative element: the treatment under consideration is differential compared to the treatment received by a similarly situated person. In this context, discrimination can be considered the opposite of equality.
Behind this apparently simple statement lies great complexity. 
As Westen~\citep{westen1982empty} demonstrated early on, the meaning of equality is lost if we do not specify what it is that makes persons or treatments "similar" in a morally relevant way.
In other words, the primary question that non-discrimination law poses is: "equal \textit{to what}?" In this section, we first provide a brief overview of how EU non-discrimination law has grappled with this question over the years, after which we discuss how discrimination is established under current EU law.

\subsection{A Brief History of EU Non-Discrimination law}
\label{sec:briefhistory}
EU law is a form of supranational law: member states of the EU transfer parts of their sovereignty to the EU, which can then legislate in specific fields.\footnote{The EU's competence is defined in Art. 2, 3, 4 and 6 TFEU \citep{tfeu}} The body of EU law comprises, among other things, the foundational treaties, secondary legislation mainly in the form of regulations and directives, and case law. While regulations apply directly within all member states, directives require member states to {transpose} their content, i.e. to implement it in their own legal system. Directives then leave member states discretion as to how the regulatory aim is to be achieved. In the field of non-discrimination law, directives reflect a minimum harmonisation approach, meaning that the law sets common minimum standards that must be achieved by all members states, but still allows individual member states to incorporate stricter measures as long as they comply with the EU treaties.

Regulations and directives are forms of statutory law: written laws that are passed by the EU legislator. It is impossible for statutory law to cover all relevant aspects of all possible cases. Consequently, to be applied in factual cases, the law needs to be interpreted by a court in a judgment. To do so, the Court of Justice of the EU takes into account the "spirit, the general scheme and the wording" of given legal provisions, including their aim as set out in the preamble and the preparatory documents, as well as previous judicial decisions that were rendered in similar cases in the past ({case law}). In the EU, a mechanism called the preliminary reference procedure allows member state courts to dialogue with the Court of Justice of the European Union (CJEU).\footnote{See Article 267 of the Treaty on the Functioning of the European Union (TFEU)~\citep{tfeu}.} Individuals are not able to access the CJEU directly, but national courts can ask questions regarding the interpretation and validity of EU law to the CJEU. After receiving the response of the CJEU, the national court then makes the final decision by implementing the CJEU's interpretation of EU law to the specific circumstances in the case at hand. \\

\noindent It is important to note that the law is not made up of static rules. In response to social advancements, new statutory law may be introduced and the interpretation of existing legal norms may change over time as new cases emerge. Over the years, EU non-discrimination law has evolved.

The first legal protection against discrimination spanning multiple European countries came with the Rome Treaty in 1957,\footnote{The Council of Europe's human rights instrument -- the European Convention on Humans Rights -- adopted in 1950 and in force since 1953, contains a prohibition against discrimination that also applies to all EU member states. The European Court of Human Rights was however only established in 1959.}
 which established the European Economic Community.\footnote{After successive treaty reforms and the entry into force of the Lisbon Treaty in 2009, its institutions were absorbed into the EU's framework.} In particular, Article 119 of the EC Treaty established equal pay for men and women.\footnote{Now Article 157 of the TFEU \citep{tfeu}. The background for this was not an agreement between the founding members to promote equality between men and women. Instead, given that at the time only France had introduced equal pay legislation, Article 119 served as an instrument for keeping a level playing field between member states in regards to expenses for the cost of labour. See also \citet{frese2020everyone}.} In 1975 and 1976, non-discrimination legislation was complemented with two directives on equality between men and women in the workplace~\citep{75/117/EEC, 76/207/EEC}. This paved the way for the Court of Justice to further elaborate non-discrimination law in subsequent years. The boundaries of EU non-discrimination law were expanded in three main directions: its application was extended to new areas, new concepts were spelled out, and new characteristics became protected against discrimination. For instance, the material scope of non-discrimination law was expanded through a broader interpretation of the notion of "pay".\footnote{See \textit{Bilka - Kaufhaus GmbH v Karin Weber von Hartz}~\citep{C-170/84} and \textit{Douglas Harvey Barber v Guardian Royal Exchange Assurance Group}~\citep{C-262/88}.} Moreover, in the landmark decision in \textit{Bilka-Kaufhaus}~\citep{C-170/84}, the Court of Justice introduced the concept of "indirect discrimination". In that case, the differential treatment was between full time and part time employees: only full time workers had access to a pension scheme as part of their employment contract. As a consequence, it was not directly covered by the wording of Article 119 which specifically guaranteed equality \textit{between women and men}. The Court, however, noted that where disproportionately more women than men work part-time, the differentiation operated by the company in granting access to the pension scheme gives rise to a discriminatory effect, in other words indirect discrimination on grounds of sex.\footnote{The Court added: "However, if the undertaking is able to show that its pay practice may be explained by objectively justified factors unrelated to any discrimination on grounds of sex there is no breach of Article 119". See \textit{Bilka - Kaufhaus GmbH v Karin Weber von Hartz, para. 30.}} In 1999, the Amsterdam Treaty entered into force and extended legal protection to other grounds of discrimination including racial or ethnic origin, religion or belief, disability, age and sexual orientation. From this point on, legislation and case law proliferated to include new regulatory territory, for instance, in the area of housing, healthcare, the consumption of goods and services and even, in limited cases, education.

Four main directives make up today's EU non-discrimination law: the Race Equality Directive 2000/43/EC~\citep{racialequalitydirective}; the Framework Equality Directive~\citep{employmentdirective}; and the gender equality Directives 2004/113/EC~\citep{2004/113/EC} and 2006/54/EC~\citep{2006/54/EC}. Additionally, primary law\footnote{There is a hierarchy of norms in EU law, according to which \textit{primary law}, which has quasi-constitutional status, prevails over \textit{secondary law} which is equivalent to legislation.} provisions include Articles 2 and 3(3) of the Treaty on European Union~\citep{teu}, Articles 8, 10, 19 and 157 of the Treaty on the Functioning of the European Union~\citep{tfeu} (the last two corresponding to ex-Article 13 EC and Article 119 EEC) as well as Articles 20, 21 and 23 of the Charter of Fundamental Rights of the EU~\citep{fundamentalrights} (the Charter), adopted in 2000 and elevated to the same status as the Treaties in 2009.

\subsection{Establishing Discrimination}
In order to understand how EU non-discrimination law operates, we need to first distinguish between the notions of direct and indirect discrimination. This distinction is key because it determines the applicable regime of justifications: direct discrimination cannot be justified except for a limited number of derogations, whereas \textit{prima facie} indirect discrimination can be justified much more widely. In other words, this technical distinction matters because it determines how the costs and burdens of inequality are distributed among decision-makers, potential victims and society at large.

Direct discrimination occurs when "one person is treated less favourably than another is, has been or would be treated in a comparable situation on grounds of" a protected characteristic~\citep{racialequalitydirective}. In other words: protected characteristics are to be excluded from any decision-making process covered by EU non-discrimination law.\footnote{In algorithmic fairness literature, direct and indirect discrimination are often equated with, respectively, disparate treatment and impact in United States law. However, an important difference between the doctrines is that while disparate treatment requires discriminatory intent, direct discrimination in EU law does not require any moral wrongdoing and will therefore apply in more cases than disparate treatment would~\citep{xenidis2019eu,adamsprassl2022directly}.} Traditionally, the doctrine of direct discrimination prescribes that "likes should be treated alike" according to the Aristotelian formula of justice as consistency, an approach often referred to as formal equality. A problem with this conceptualisation of equality is that it is unable to redress more complex forms of injustice such as proxy discrimination and structural inequality. For example, a rule banning all individuals shorter than 1,70m from applying to jobs with the police essentially excludes a large majority of women. Yet the selection does not depend explicitly on the sex or gender of candidates, and therefore it does not amount to direct discrimination on grounds of sex as confirmed by the CJEU in \textit{Kalliri}~\citep{C-409/16}.

As explained in the previous section, to complement the legal protection of equality, the Court of Justice has adopted the doctrine of {indirect discrimination}, which, in certain situations, forbids treating those who are unalike in a like manner. Specifically, indirect discrimination occurs where "an apparently neutral provision, criterion or practice would put persons of a protected group at a particular disadvantage compared with other persons, unless that provision, criterion or practice is objectively justified by a legitimate aim and the means of achieving that aim are appropriate and necessary"~\citep{racialequalitydirective}. 
This asymmetrical conception of equality encapsulates the second part of the Aristotelian formula and forbids applying the same rule to legal subjects who are positioned differently. Our example above, concerning the application of the same height requirement to male and female candidates, falls within the concept of indirect discrimination~\citep{C-409/16}. The ban on indirect discrimination has often been described as guaranteeing a substantive form of equality because it creates an obligation to accommodate legally protected differences (for instance height difference resulting from one's sex) and associated lifestyles (for instance protecting certain religious holidays). Since indirect discrimination focuses on the disadvantageous effects of given rules and practices rather than the inclusion of protected characteristics in given decisions, it allows addressing proxy discrimination that impacts protected groups. To some extent, this creates an obligation for decision-makers to account for the unjust \textit{status quo} that prevails in society. For example, the gender pay gap is a well-known form of institutionalised discrimination. The practice of using newly recruited employees' past salaries to decide on their new pay in salary negotiations could be regarded as indirect discrimination on grounds of sex, because it tends to perpetuate the gender pay gap.

From the definitions of direct and indirect discrimination, we can identify four main elements in a discrimination case. 

\paragraph{"On grounds of"...}
To determine whether the case is one of direct or indirect discrimination, it is necessary to assess whether a decision was taken "on grounds of" a protected characteristic. When a protected characteristic is explicitly used as a basis for a decision, that decision falls under the notion of direct discrimination. In some cases, using a proxy that is "inseparably linked" to a protected ground (e.g. pregnancy and sex) will amount to direct discrimination~\citep{C-177/88}. By contrast, if a decision creates a disadvantage to a protected group albeit not targeting that group, it falls within the notion of indirect discrimination.

\paragraph{..."a protected characteristic" in an area covered by EU law (personal and material scope)...}
Protected characteristics vary across sectors. The widest protection against discrimination can be found in relation to employment, where discrimination is banned in relation to racial or ethnic origin, sex or gender, religion or belief, disability, age and sexual orientation~\citep{racialequalitydirective, employmentdirective, 2006/54/EC}. In relation to access to goods and services, only racial or ethnic origin and sex or gender are protected characteristics. Although a major concern from a social or moral point of view is that algorithmic systems operate differently based on people's income or socio-economic background, this form of disadvantage does not fall within the scope of protection offered by EU secondary law.
In addition, while discriminatory effects may occur at the intersection of two or more vectors of disadvantage (for example race and gender or age and sexual orientation),~\citep{crenshaw1990mapping, hoffmann2019fairness}, the CJEU has so far failed to recognise intersectional discrimination explicitly.\footnote{It could be argued that the Court has nevertheless addressed combined discrimination implicitly in cases such as \textit{Odar} or \textit{Bedi}, which combined disadvantage based on age and disability.~\citep{C-152/11, C-312/17}.} For example, in \textit{Parris}~\citep{C-443/15} the Court found that no "combined" discrimination on grounds of sexual orientation and age could exist where discrimination could not be proven on each ground taken separately~\citep{atrey2018illuminating, xenidis2018multiple}.

Directive 2004/113/EC~\citep{2004/113/EC} includes some exceptions for the ban on gender discrimination, namely in relation to advertisement and the media as well as education. By contrast, discrimination on grounds of racial or ethnic origin is prohibited in relation to education. Furthermore, Article 21(1) of the Charter prohibits discrimination based on a greater number of grounds than secondary law, including but not limited to sex, race, colour, ethnic or social origin, genetic features, language, religion or belief, political or any other opinion, membership of a national minority, property, birth, disability, age or sexual orientation. Article 21(1) of the Charter and the general principle of equality are both horizontally and vertically directly applicable (i.e. they have direct effect in relations between public and private parties and between private parties themselves)~\citep{C-144/04,C-414/16}.\footnote{In principle, the Charter is only directly applicable in vertical relationships between public authorities and private parties. However, the Court has carved out horizontal direct effects in relation to several articles including Art. 21(1) on non-discrimination in C-414/16 \textit{Egenberger}~\citep{C-414/16} as confirmed in C-68/17 \textit{IR v JQ}~\citep{C-68/17}, and Art. 31 on annual leave in Joined Cases C-569/16 and C-570/16 \textit{Bauer}~\citep{C-569/16}.} By contrast, directives are only vertically directly applicable, meaning that their provisions only apply directly between a public and a private party.\footnote{Direct effects arise only in relation to provisions that are precise, clear and unconditional, see C-26/62 \textit{Van Gend en Loos}~\citep{C-26/62}.} However, national law transposing directives could in and of itself create horizontal effects.

\paragraph{...where there is evidence for "less favourable treatment" or "particular disadvantage"...}
To establish a case of discrimination, an applicant first needs to bring \textit{prima facie} evidence, i.e. sufficient evidence for a rebuttable presumption of discrimination to be established by the judge. Evidence of \textit{prima facie} \textit{direct} discrimination could include, for instance, information about another group or individual of a different protected group being treated more favourably. If such a comparator does not exist, EU law allows applicants to construct a hypothetical comparator. Evidence of \textit{prima facie} \textit{indirect} discrimination involves raising a reasonable suspicion that a given disadvantage affects a protected group. This could, but does not have to, involve statistics.\footnote{By contrast with US law which relies a lot on statistical evidence, evidence in EU law is much more contextual and hardly relies on statistical comparisons.} If \textit{prima facie} discrimination is established, the burden of proving that discrimination has not occurred shifts to the defendant.

\paragraph{...unless there is an "objective justification"}
While direct discrimination is not justifiable in principle (except for a few exceptions provided for by the law), the indirect discrimination doctrine allows for a \textit{prima facie} discriminatory measure to be "objectively justified" where it fulfils a legitimate aim and passes the so-called proportionality test. The law does not provide concrete guidelines on whether the means to achieve a legitimate aim are necessary and proportionate. Due to the large variety yet small number of cases, the proportionality test cannot be settled in advance based on previous case law. One rule that stands out is that if the same legitimate aim can be achieved through less discriminatory alternatives, those must be used~\citep{tobler2005indirect}. Other than that, however, objective justifications are judged on a case-by-case basis, depending on the significance of the harm and the legitimacy of the aim.

\section{Algorithmic Unfairness through the Lens of Non-Discrimination Law}
\label{sec:comparison}
Over the past years, several incidents have raised concerns regarding bias and unfairness of algorithms and, in particular, AI systems. When used in automated decision-making, AI systems have the ability to produce fairness-related harms systematically and at a large scale. Moreover, while discrimination by human actors can to some extent be signalled to victims through behaviour or past experiences, discrimination by algorithmic systems typically remains largely invisible. In light of the increased use of machine learning systems, it has thus become a pressing question to which extent algorithmic unfairness can be seen as discrimination under EU law~\citep{xenidis2019eu}. In this section, we analyse several seminal examples from algorithmic fairness literature through the lens of EU non-discrimination law.

\subsection{Dutch Childcare Benefits Scandal}
We start our analysis with a case related to the explicit use of a sensitive feature in a machine learning model, which is often assumed to be unlawful.
In January 2021, the Dutch government resigned over a scandal involving false fraud allegations made by the Tax and Customs Administration in the distribution of childcare benefits. In particular, over the course of several years, the administration had used a risk assessment algorithm that explicitly included Dutch citizenship as one of the risk factors.\footnote{While our analysis focuses on the used risk assessment algorithm, we would like to emphasise that the scope of the scandal was much broader, involving the complete working procedure of the Tax and Customs Administration.}
To determine whether this is a case of unlawful discrimination under EU law, we first need to determine whether it falls within the material and personal scope of EU non-discrimination law.
This particular case involved a public body and, if the case fell within the scope of EU law, Article 21(1) of the Charter, which prohibits discrimination on a non-exhaustive list of grounds including membership of a national minority, could apply. Indeed, the Dutch Data Protection Authority (DPA) established that the use of nationality as a factor in the risk classification model is considered discriminatory processing of data on the basis of, amongst others, Art. 21 of the Charter, and therefore illegitimate given the principle of fairness in Article 5 of the GDPR~\citep{dpa2021fine}.\footnote{Note that Article 51(1) restricts the scope of application of the Charter only to situations where "Member States [...] are implementing Union law". In this case, the GDPR can provide the necessary link to EU law to the extent that public authorities are implementing EU data protection legislation when processing data. Note that the case might also be framed as one of discrimination on grounds of ethnicity, in which case the Race Equality Directive 2000/43/EC might be applicable. The Court has dealt with similar issues in cases such as C-668/15 \textit{Jyske Finans} and C-457/17 \textit{Maniero}.} In particular, the DPA explained that incorporating nationality as a factor in the risk classification model could result in higher risk scores for applicants who are not Dutch citizens compared to applicants with a Dutch nationality~\citep{dpa2021report}. This increased the probability of higher scrutiny through manual processing of the application by an employee of the tax administration, which the DPA considered a particular disadvantage.\footnote{At first glance, this seems like a clear case of direct discrimination: the algorithm explicitly included nationality as a factor in decision-making. Instead, however, the DPA analysed the case through the lens of indirect discrimination: nationality by itself is insufficient to determine whether the applicant is eligible for childcare benefit, as it is also relevant whether an applicant is registered in a Dutch municipality or is a lawful resident in the Netherlands. Thus, the DPA explained, the tax administration could have used a risk factor with less potential for discriminatory effect, such as: "applicant possesses Dutch nationality, \textit{or} EU nationality and is registered in a Dutch municipality, \textit{or} a non-EU nationality and has a valid residence permit".}

However, even in cases of (in hindsight) obvious potential for discriminatory treatment, establishing \textit{prima facie} evidence can prove very difficult -- especially in the context of unintelligible or inaccessible algorithms. In case of the childcare benefits scandal, parents were wrongly accused over the course of a decade and the full scale of the scandal only became clear after several years of investigation. Notoriously, parents who requested access to their files received documents with pages and pages of redacted text~\citep{nos2019zwartgelakt}. In a situation like this, the case law of the CJEU shows that the absence of transparency or information can contribute to contextual evidence with a view to triggering a shift of the burden of proof~\citep{C-109/88}. Yet, when algorithmic systems are embedded into opaque decision-making processes, an individual is unlikely to become aware that discrimination has occurred at all. Therefore, legal claims of discrimination might not even arise without adequate support. This raises questions regarding the protection that equality law, which is designed to protect against discrimination by humans, offers in cases of algorithmic discrimination.

\subsection{Amazon's Recruitment Algorithm}
A commonly cited example of algorithmic bias is a resume selection algorithm that was under development at Amazon in 2017~\citep{dastin2018amazon}. As it turned out, the algorithm penalised words that indicated the applicant's gender, such as participation in the women's chess team or attending an all-woman's college. It is important to note that Amazon's hiring algorithm was not necessarily less accurate for women compared to men. Instead, the main culprit for the disparity was unequal hiring rates: in the past, the company had primarily hired men for technical roles. An important question is why these hiring rates differed. We can identify at least two potential reasons: either the data is a biased measurement of reality or reality is biased.\footnote{While this may seem to suggest that algorithmic unfairness is primarily related to biases in data sets, we would like to emphasise that algorithmic bias is not merely a problem of \textit{"bias in = bias out"}. Data sets do not simply exist, they are constructed. Considering a backdrop of historical injustice and structures of oppression, the social processes that produced these data sets require critical attention. Having said that, the causes of fairness-related harms induced by algorithmic systems can -- in both subtle and obvious ways -- be different from harms induced by human actors. Therefore, we believe an increased understanding of the different ways in which algorithmic systems can cause harm is critical for their mitigation.}
First, we might be looking at a case of measurement bias: historical hiring decisions are incomplete measurements of actual employee quality. When measurement bias is associated with a sensitive characteristic, in this case gender, the model is likely to replicate the pattern which can result in an unfair allocation of jobs~\citep{jacobs2021measurement}. In other words, the sensitive characteristic is implicitly included as a factor in decision-making. This type of unfairness speaks to the exclusionary function of formal equality: protected characteristics should be excluded from decision-making. Second, gender disparities in hiring rates could in part be explained by disparities in behaviour caused by factors related to structural inequality. For example, women may have been systematically discouraged from pursuing technical roles, resulting in fewer suitable candidates. From this perspective, the wrongness of Amazon's hiring algorithm can best be considered through the lens of substantive equality.

How would such a case of algorithmic unfairness be captured by EU discrimination law? According to Amazon, the algorithm was never actually used. For the sake of our argument, however, let's assume that the algorithm was deployed in the EU. Employment discrimination on the basis of gender clearly falls within the material scope of non-discrimination law. While gender is not used directly as a factor by the algorithm, penalising applicants on the basis of characteristics highly associated with the applicant's gender can be seen as a form of proxy discrimination that would either fall under the indirect discrimination doctrine or, in line with the Court of Justice's jurisprudence in \textit{Dekker}~\citep{C-177/88} under the direct discrimination doctrine if the decision criteria used are "inextricably linked" with sex or gender. As argued by \citet{adamsprassl2022directly}, we may wonder to what extent attendance of an all woman's college can be seen as an "apparently neutral criterion" that is not inseparably linked to gender. As mentioned above, the distinction between direct and indirect discrimination is key because it determines whether observed disparities can be justified, and ultimately who is responsible for internalising the costs of social inequality.

From a conceptual perspective, predicting how the Court of Justice would legally qualify the Amazon recruitment algorithm raises at least two issues. First, the Court of Justice has not always consistently distinguished between direct and indirect discrimination. For instance, in \textit{Dekker}~\citep{C-177/88}, the Court ruled that discrimination on grounds of pregnancy amounted to direct discrimination on grounds of sex because of the "inextricable" link that exists between pregnancy and sex. As a result, even where the protected characteristic itself was not used as a basis for a decision, using a proxy that is "inseparably linked" to it amounts to direct and not indirect discrimination. At the same time, it is unclear which proxies will be regarded as "inseparably linked" to protected characteristics. In \textit{Jyske Finans}~\citep{C-668/15}, the CJEU did not consider that the practice of a credit institution to subject an EU citizen to an additional identity check when born outside the EU amounted to direct discrimination on grounds of racial or ethnic origin. The CJEU did not deem the link between someone's country of birth and ethnic origin "inseparable".\footnote{To answer the question of the nature of the link, it is first necessary to define what ethnic origin is and in relation to which group(s), which is a delicate question. Here the differentiation was between EU- and non-EU-born citizens.} In sum, the boundary between direct and indirect discrimination is contested and the Court has not always been consistent in distinguishing both notions or in defining what "on grounds of" a protected characteristic means.\footnote{It has also been argued that, from a moral point of view, direct and indirect discrimination capture the same harm~\citep{moreau2020faces}.} 

Second, part of the problem of distinguishing between direct and indirect discrimination is linked to the difficulty of defining what a protected characteristic is. The answer to this question directly depends on the choice of comparator made by the Court.\footnote{As argued by Westen, the comparator simultaneously defines the normative baseline of discrimination law, that is the desirable level of equality in a given situation\citep{westen1982empty}.} For instance, in the context of neutral dress codes imposed by employers on their employees, whether or not discrimination is deemed direct or indirect heavily depends on which comparator is chosen. If religious and non-religious employees are compared, it appears that not all religious employees are disadvantaged by the rule. This seems to exclude direct discrimination. However, if employees whose religion mandates wearing religious clothing and employees whose (absence of) religion does not are compared, this reveals that a well-defined group is exclusively disadvantaged by the rule~\citep{cloots2018safe, sharpston2022shadow}, because the rule is more compatible with some religious practices than others. In fact, the divide between direct and indirect discrimination has been extensively discussed by commentators in the context of the so-called headscarf cases. In its \textit{Achbita}~\citep{C-157/15} and \textit{Wabe}~\citep{C-804/18} decisions, the Court has been criticised for failing to treat facially neutral dress codes as a form of direct discrimination on grounds of religion (and gender)~\citep{mulder2022religious}.\footnote{Note that the Court distinguishes the situation in \textit{Wabe} from that in \textit{Müller}.} As former Advocate General Sharpston stated, "`neutrality' that in reality predictably denies employment opportunities to particular, very clearly identifiable, minority groups is false neutrality" and should thus not fall within the scope of indirect discrimination~\citep{sharpston2022shadow}.

Given the Court's problematic approach to the distinction between direct and indirect discrimination, there is a risk that the Court could treat cases of algorithmic unfairness such as Amazon's recruitment algorithm from the perspective of indirect discrimination. This would raise two further issues. First of all, the notion of "particular disadvantage" inherent in indirect discrimination is particularly vague, which makes it difficult both to assess compliance and to provide evidence for \textit{prima facie} discrimination. For example, in \textit{Kalliri}~\citep[][para. 31]{C-409/16}, the Court found evidence of \textit{prima facie} discrimination because the height requirement of 1,70m "work[ed] to the disadvantage of \textit{far more} women than men". The existence of a particular disadvantage is only assessed by the Court contextually. In \textit{Seymour-Smith}~\citep{C-167/97} the Court considered that statistics showing that 77.4\% of the men and 68.9\% of the women in the workforce were able to meet the two-year employment requirement needed to obtain compensation for dismissal "d[id] not appear, on the face of it, to show that a considerably smaller percentage of women than men is able to fulfil the requirement" \citep{wachter2021fairness}. However, there is no consistent use of statistics by the Court. The normative principles guiding this assessment and the thresholds operated by the Court of Justice often remain implicit.\footnote{The Court sometimes explicitly reasons in terms of observable structural inequalities (e.g. the caregiver/breadwinner divide, the gender pay gap, the gender pension gap, stereotypes, etc.), but often without quantifying lawful and unlawful imbalances.} We can see those elements emerge in a few cases such as \textit{YS v NK}~\citep{C-223/19}, which concerned a claim of indirect discrimination on grounds of sex, age and property. The Advocate General dismissed the applicant's argument that an austerity measure cutting a type of large pensions in use in the 1990s amounted to a particular disadvantage against older men. If the comparison test showed that men were affected more by the measure than women in absolute terms, she reasoned that it would “at most [be] linked to an already existing state of inequality”. In other terms, gender segregation on the labour market in the 1990s, the current gender pay gap and the gender pension gap would explain any apparent impact on older men: any “predominant impact on men would in all likelihood have to be solely attributed to the fact that men, on average, still earn more than women and are over-represented in management positions”.~\citep[][para. 64 and 76]{C-223/19} This case reveals the normative principle underpinning the Court's assessment of a "particular disadvantage": the lens of indirect discrimination should capture the unjustified reinforcement of inequalities as opposed to mere punctual "unbalances". Hence, rather than targeting a precise threshold, probing legal compliance in situations of algorithmic unfairness requires reflecting on the implications of a given imbalance in terms of structural inequality.

Second, the indirect discrimination doctrine allows for an objective justification. If Amazon's hiring algorithm is interpreted as indirect discrimination, the accuracy of the algorithm on a test set may be deemed an acceptable justification in court~\citep{adamsprassl2022directly}.\footnote{It has even been argued that "[i]n most scenarios, indirect discrimination produced by ML systems will pass the proportionality test of the CJEU"~\citep{martinez2022discriminatory}.}  Without access to information regarding the data collection procedure and machine learning process, it is difficult for applicants to prove whether accuracy -- as indicated by the alleged offender -- is a good reflection of effectiveness in practice. However, in cases of outcomes tainted by measurement bias, accuracy on \textit{observed} data is an inadequate measurement of the true effectiveness of the model. Moreover, accuracy in a test environment may not generalise to accuracy of the algorithm after deployment, particularly in cases of out-of-sample predictions (i.e. the model is used under circumstances different from the one it was trained on) or concept drift (i.e. the data distribution evolves over time).\footnote{In the common position adopted by the Council of the EU in November 2022, Art. 10(3) of the current proposal for an EU AI Act stipulates that "[t]raining, validation and testing data sets shall be relevant, representative, and to the best extent possible, free of errors and complete".} Importantly for computer scientists thinking about how to translate legal norms to ensure compliance, the normative principle underpinning the Advocate General's reasoning in \textit{YS v NK}, i.e. substantive equality, can be used to shape the proportionality test. As confirmed by AG Kokott, "the existing economic inequality between the sexes is not exacerbated further in the present case" so "the requirements regarding the justification of any indirect discrimination are correspondingly lower". In other words, even though \textit{prima facie} a particular disadvantage arises punctually, it can be justified if it does not generate or reinforce structural inequalities.\footnote{It is important to note, however, that the Court has also been criticised for an inconsistent approach to the normative underpinnings of the doctrine of indirect discrimination, i.e. it does not always consistently approach indirect discrimination from the perspective of substantive equality.} This is an important indicator for assessing legal compliance.

\subsection{Gender Shades}
In their seminal work "Gender Shades", \citet{buolamwini2018gender} found that several commercial facial recognition systems intended to identify a person's gender failed disproportionately for darker-skinned women, particularly compared to faces of lighter-skinned men. There are many reasons why the predictive performance of a machine learning system differs across groups, including the use of features that are not equally predictive across groups and the use of a machine learning algorithm that is unable to adequately capture the data distributions of minority groups. In the case of Gender Shades, the primary culprit was the under-representation of darker-skinned women in facial recognition data sets. This type of bias can be particularly problematic when the data distribution of majority groups differs substantially from the data distribution of minority groups.\footnote{This can itself be the result of structural inequality, e.g. unequal access to a given set of jobs, educational opportunities, housing options, etc.\citep{hellman2021big}.}

Again, we first need to consider whether the problem at stake falls within the material scope of EU discrimination law, which itself depends on the sector in which the facial recognition system is used. For example, if facial recognition is required to gain access to particular goods or services (with the exception of advertisement, education and the media in relation to gender-based discrimination), disparate misclassification rates in relation to gender or skin colour lead to denying access to protected groups fall within the material scope of Directives 2004/113/EC~\citep{2004/113/EC} and Directive 2000/43/EC~\citep{racialequalitydirective}.\footnote{In our example, the directives cover goods and services available to the public that are sold both by private and public parties. Furthermore, the broad protection against discrimination anchored in Art. 21 of the Charter applies in relation to public bodies when they are implementing EU law.} As race and gender are not used directly as input factors in the algorithms, a case like this might fall within the indirect discrimination doctrine.\footnote{\citet{adamsprassl2022directly} have pointed out the limitation of usual interpretations of "because of" in direct discrimination, which is primarily designed to combat human discrimination. Even when protected characteristics are not used as factors at the point of decision-making, it is hard to view disparate predictive performance in facial recognition as not causally dependent on race and gender.} This would open up the possibility of an objective justification.

For example, in 2022, a Dutch student filed a complaint against her university, stating that the face recognition check included in fraud detection software used during online exams, often failed -- seemingly due to the student's dark skin colour. In an interim judgement, the Netherlands Institute for Human Rights states that the disadvantage experienced by the student, together with scientific research pointing towards disparate performance of face recognition algorithms, provide \textit{prima facie} evidence for indirect discrimination in relation to race \citep{crm2022interimjudgment}\footnote{The Institute specifically refers to Article 7(1)c of the Dutch AWGB (Algemene Wet Gelijke Behandeling), which prohibits discrimination based on race (which should be interpreted broadly to also include skin colour) with regard to access to goods and services by institutions in the field of education.} and shifting the burden of proof to the university to prove the law was not violated.

Furthermore, the case of facial recognition software provides an interesting case study for interrogating the boundaries of EU non-discrimination law. Would a particular disadvantage arising from the disparate \textit{quality} of goods and services, for instance, face recognition, in relation to gender or race fall within the ban on discrimination? Arguably, there is a case for EU non-discrimination law in the area of goods and services to be applied to disparate product safety and performance across demographic groups. For example, could the exclusive use of male crash dummies to test cars be captured by the Gender Directive 2000/43/EC on goods and services, since it results in higher risks of injury for female occupants \citep{linder2019road}? Even though the case law in this area is scarce and does not provide for immediate analogies (see e.g. C-236/09 \textit{Test-Achats}~\citep{C-236/09}), the scholarship in this area points towards the applicability of EU non-discrimination law~\citep[][p. 94]{caraccioloditorella2022}.\footnote{In \textit{Test-Achats}, the Court struck down the use of gender by insurance companies as an actuarial factor to assess risks and calculate the price of insurance policies.} In addition, the Court's inclusion of the notion of `access' within the scope of protection of EU law in \textit{Maniero}, a case concerning the award of educational scholarships, points towards the applicability of EU non-discrimination law to harms related to disparate quality of service. In that case, the Court indicated that "there can be no education without the possibility to access it" and that "the directive’s objective, which is to combat discrimination in education, could not be achieved if discrimination were allowed at the access to education stage"~\citep[][para. 37]{C-457/17}. In addition, the Advocate General in \textit{Maniero} endorsed a broad interpretation of the notion of `access': "access to education has many component parts. It could be physical access to a building; imposing a \textit{numerus clausus} system to keep student numbers controlled; the ability to borrow or purchase books; the ability to pay for living expenses (amongst many others)"~\citep[][para. 33]{C-457/17opinion}. By analogy, the disparate quality or performance of algorithmic systems for protected groups could be understood as affecting their access to goods and services in a discriminatory manner. In such an extensive interpretation of non-discrimination guarantees, biased systems like the face recognition tools in our example could potentially fall under EU non-discrimination law regardless of whether they condition access to other goods and services.\footnote{A crucial question would then be what disparity rates are considered to amount to discrimination.}

Finally, an important characteristic of the Gender Shades study was the emphasis on intersectional concerns: while facial recognition systems generally performed worse for women and people of colour, the disparity was the greatest for darker-skinned women. As mentioned above, EU law does not prevent the CJEU from considering intersectional discrimination, but the Court has so far failed to properly engage with this issue.\\

\noindent From these examples, we can see that EU non-discrimination law is in principle suited to deal with types of algorithmic unfairness that closely resemble human discrimination. However, reasoning by analogy to apply legal norms and principles to cases of algorithmic unfairness reveals grey areas and inconsistencies in the Court's approach to discrimination. Some of these gaps could be filled via teleological interpretation of EU discrimination law in the digital context, for example in cases of disparate predictive performance, but this also opens up difficult normative questions. Moreover, the unintelligibility of prediction-generating mechanisms and lack of transparency regarding important design choices of AI systems make it difficult for applicants to provide \textit{prima facie} evidence to even start court proceedings. From a legal compliance perspective, since the CJEU rarely relies on statistical evidence in its judgments, it is difficult to derive general, abstract or readily transferable rules of thumb regarding requirements for thresholds, proportionality or justification from the highly particularised case law of the Court.

\section{The Problem of Emptiness}
\label{sec:normative}
In response to concerns regarding algorithmic bias and unfairness, computer scientists have proposed several fairness metrics and fairness-aware machine learning (fair-ml) algorithms that are designed to measure and mitigate fairness-related harm. A straightforward question, then, is which fairness metric AI practitioners should choose and what value it should take in order to be compliant with the law. From the examples in the previous section, it is clear that EU non-discrimination law does not provide us with explicit rules that must be upheld. Instead, the court is granted judicial discretion that allows it to make normative decisions based on the specifics of an individual case -- an approach \citet{wachter2021fairness} refer to as "contextual equality". To better understand the applicability of fairness metrics and the algorithms that optimise for them, we must therefore consider their normative underpinnings.

\subsection{Emptiness in Fairness Metrics}
A common denominator of algorithmic fairness metrics is equality - be it in the form of a particular distribution of predictions in the case of group fairness metrics, approximately equal treatment in the case of individual fairness, and equal counterfactual outcomes in case of counterfactual fairness. The choice of fairness metric, then, boils down to a question that greatly resembles the primary question of non-discrimination law: \textit{what} should be equal? 
The most prominent fairness metrics in algorithmic fairness literature concern the classification scenario, where we can distinguish two main lines of work: group fairness and individual fairness.

\textit{Group fairness} metrics aim to capture the extent to which particular group statistics are equal across sensitive groups. Similar to protected characteristics in non-discrimination law, sensitive features are intended to represent group membership of some socially salient group. Numerous group fairness metrics have been proposed in algorithmic fairness literature, which can be differentiated primarily in terms of which group statistic is compared.
Arguably the strongest requirement of equality is set by \textit{demographic parity}\footnote{Taking inspiration from the US disparate impact doctrine, demographic parity is sometimes referred to as disparate impact~\citep[e.g.][]{feldman2014DisparateImpact}, which several scholars have argued to be overly reductive~\citep[e.g.][]{watkins2022fourfifths}.}, which requires the proportion of positive predictions (e.g. the selection rate in hiring) to be equal between groups. For example, in case of Amazon's recruitment algorithm, a positive prediction relates to a benefit (a job interview) and demographic parity essentially requires that receiving the benefit should be independent of sensitive group membership -- even if observed data suggests otherwise. By choosing demographic parity as a fairness metric, we thus implicitly assume that whether an individual is deserving of the benefit does not depend on their observed ground truth class. This can be an empirical assumption, e.g. because we believe that observed data is subject to measurement bias, or it can be a more explicit normative assumption, e.g. that the observed ground truth class is affected by historical injustice that we do not wish to replicate~\citep{hertweck2021moral}.
Contrarily, the group fairness metric \textit{equalised odds}~\citep{hardt2016equality} considers an individual's ground truth class a factor that can justify existing disparities in the distribution of predictions. Specifically, this metric considers equality of false positive rates (e.g. the proportion of healthy individuals that are falsely predicted to have a disease) and false negative rates (e.g. the proportion of sick individuals that are predicted to be healthy). In the case of the distribution of a benefit, the use of this metric thus reveals a specific normative assumption: the status quo is acceptable \citep{wachter2020bias}.
A third commonly cited metric, \textit{equal calibration}, requires that predicted scores are equally well calibrated across groups. A model is considered to be well calibrated if the output of the model (i.e. predicted scores) corresponds to the probability of belonging to the positive class.\footnote{Calibration is particularly relevant when predicted scores are used as input in decision-making, as a decision threshold for calibrated scores can be directly interpreted in term of different misclassification costs. For example, if a calibrated confidence score is used for suggesting a specific treatment in clinical decision-making, a decision threshold of 0.1 means that we accept up to 9 false positives (i.e., unnecessary treatments) for each true positive.} For example, a model is calibrated if out of all instances that receive a predicted score of 0.7, the proportion of instances that actually belongs to the positive class is also 0.7. Essentially, equal calibration requires that the meaning of predicted scores is equal across groups~\citep{jacobs2021measurement}: receiving a score of 0.7 corresponds to a probability of 0.7, irrespective of sensitive group membership.
In contrast to demographic parity and equalised odds, equal calibration cannot be readily interpreted as a particular distribution of burdens and benefits and instead relates more to \textit{beliefs} about (groups of) individuals~\citep{hellman2020measuring}.

Where group fairness metrics primarily consider fairness from the perspective of groups of people, notions of individual and counterfactual fairness are primarily concerned with the perspective of the individual. 
\textit{Counterfactual fairness} metrics consider fairness from an explicit causal modelling perspective~\citep{kusner2017counterfactual}. An elaborate explanation of causal inference is outside of the scope of this paper -- it suffices to know that empirical assumptions regarding causal relationships between (sensitive) features and outcome variables are modelled in a causal graph. Counterfactual fairness, then, considers the question: given what we know about this individual, how would the model's prediction change, had they belonged to a different sensitive group? If the prediction changes, the model does not satisfy counterfactual fairness. The underlying normative assumption, then, is that factors that are causally related to sensitive group membership should not impact the outcome.

Normative assumptions become less explicit when we consider metrics that allow the user to specify characteristics that may justify observed disparities. At the extreme, \textit{individual fairness} requires that "similar people are treated similarly" \citep{dwork2012awareness}. Here, similarity is measured through a quantitative similarity metric, usually based on the input features. Essentially, all normative assumptions are therefore captured in the choice of similarity metric \citep{binns2020apparent}. Inspired by the notion of "objective justification" in the indirect discrimination doctrine, some variations of fairness metrics, such as \textit{conditional demographic parity}~\citep{kamiran2013quantifying,wachter2021fairness} and path-specific counterfactual fairness~\citep{chiappa2019path}, allow further conditioning on specific characteristics that are deemed justifiable factors in decision-making irrespective of their (causal) relationship to a sensitive characteristics. For example, in college admissions, we may want to account for varying levels of competitiveness across programs. That is, instead of measuring whether overall admission rates are equal for female and male applicants, we measure equality of selection rates within each program separately.

\subsection{Emptiness in Fairness-Aware Machine Learning}
In addition to fairness metrics, much work in algorithmic fairness research has centred around technical interventions purporting to mitigate unfairness, which we will refer to as fairness-aware machine learning (fair-ml) techniques. A typical approach is to formulate the problem as an optimisation task, where predictive performance is optimised subject to a fairness constraint.\footnote{We refer the interested reader to \citet{caton2020fairness} for a comprehensive overview of existing techniques.}
Fair-ml techniques are commonly categorised into three groups. \textit{Pre-processing} approaches modify the data used to train the ML model. Most pre-processing techniques aim at ensuring that the sensitive feature and target variable are statistically independent. For example, the output label (e.g. "hired" or "not hired") of (some) instances in the training data set may be changed according to an algorithmic heuristic. In contrast, \textit{in-processing} techniques incorporate fairness constraints directly into the machine learning process. For example, instead of optimising solely for misclassification errors, we can include a penalty in the objective function that quantifies to what extent the model deviates from a particular fairness constraint. Finally, \textit{post-processing} algorithms account for fairness after a model has been trained, including direct adjustments to the model parameters or adjustments to the predictions of the model. For example, to account for disparate hiring rates across genders, we may adjust the decision threshold for one group (e.g. male applicants) such that the proportion of hired individuals is equal.

Some fair-ml algorithms are explicit regarding the underlying empirical and normative assumptions. For example, the massaging technique introduced by \citet{kamiran2013quantifying} relies on the assumption that discrimination is most likely to occur to individuals close to the decision boundary of a classifier. Consequently, the algorithm relabels instances considered to be border cases such that the base rates are equal across sensitive groups. Similarly, the reject-option classification~\citet{kamiran2012decision} approach essentially applies a different decision threshold across sensitive groups, centred around the original decision threshold. As such, these techniques can be interpreted to counteract a specific form of measurement bias in which particular groups receive systematically lower scores. However, despite often referred to as "de-biasing" techniques, many fair-ml techniques do not explicitly counteract biases that lie at the root of fairness-related harm, but instead optimise directly for a given fairness constraint. For example, pre-processing techniques intended to learn new representations of the data~\citep{zemel2013learning} and constrained learning techniques cannot be readily interpreted as particular decision-making policies. Instead, these techniques take an effects-based approach, assuming that as long as a fairness constraint is satisfied, biases have been counteracted. This can be problematic, especially considering the under-specification of fairness metrics from a normative standpoint. Consequently, simply enforcing a metric by means of a fair-ml technique can have various undesirable consequences. For example, some algorithms enforce equality by reducing benefits for the advantaged group, rather than increasing benefits for the disadvantaged group~\citep{weerts2022does}. Notably, such a levelling-down approach is contrary to the case law of the Court of Justice, which indicated in \textit{Milkova} that redressing discrimination requires "granting to persons within the disadvantaged category the same advantages as those enjoyed by persons within the favoured category” where there is “a valid point of reference” and “as long as measures reinstating equal treatment have not been adopted”~\citep[][para. 32]{C-406/15}.

\subsection{Emptiness in the Law}
Many of the aforementioned fairness metrics are incompatible with each other. In particular, when base rates (i.e. the proportions of positives) differ between groups, any combination of demographic parity, equalised odds, and equal calibration cannot be satisfied simultaneously~\citep{kleinberg2016inherent,chouldechova2017fair}. Additionally, when all input features are incorporated in a similarity metric, individual fairness is typically at odds with demographic parity~\citep{dwork2012awareness}. Given the vastly different empirical and normative assumptions of these metrics, this should not come as a surprise. In particular, different metrics make different assumptions regarding the characteristics that can justify disparities. This brings us back to the problem of emptiness inherent in the principle of equality: what factors should or should not play a part in decision-making? And what normative baselines should be used to assess the right equality standard, the right amount of benefit received or the right quality of treatment? In the next paragraphs, we seek guidance in the legal reasoning of the CJEU. 

In some cases, case law provides us with such guidance. Considering a measure withdrawing benefits from an advantaged group to ensure equality with a disadvantaged group, the Court has been clear. For example, in \textit{Cresco}~\citep{C-193/17}, a private employer applied a piece of discriminatory legislation concerning religious holidays. The Court of Justice ruled that it could not simply withdraw the benefit from the "advantaged" group of workers to reinstate equality, but rather that it had to extend the benefit to all workers across the protected group (religion). This shows that equal treatment on the face of it is insufficient and that the question "equal to what" was answered by the Court by pointing at the most advantaged group.\footnote{For other examples of the "levelling up" approach see \cite{xenidis2021polysemy}.} 

Next, we can consider fair-ml approaches that set group-specific decision thresholds by analogy with the case law of the Court on so-called positive action measures, and in particular quotas. The Court of Justice has been particularly strict when assessing the lawfulness of quotas. In \textit{Kalanke} and \textit{Marschall}, for example, the Court only allowed \textit{flexible} as opposed to strict, unconditional, automatic or absolute quotas~\citep{C-450/93, C-409/95}.\footnote{In \textit{Marschall}, the Court allowed the quota because it contained a so-called "saving clause" "to the effect that women are not to be given priority in promotion if reasons specific to an individual male candidate tilt the balance in his favour"~\citep[][para. 24]{C-409/95}.} In addition, EU equality law does not \textit{require} but only \textit{allows} positive action measures. Therefore, ensuring the lawfulness of post-processing techniques might amount to walking a tightrope.

Unfortunately, the law is not always as clear. As demonstrated by \citet{schauer2018treating}, the question of similarity central to judicial precedent and to the comparative heuristics that underpin the Court's discrimination test is not as such an ontological question of similarity, but instead revolves around what the Court \textit{deems} similar. The CJEU has not been explicit regarding the normative framework that is used to determine what makes two cases similar, resulting in inconsistencies.\footnote{The Court has, however, constantly made clear that two cases do not need to be similar in absolute terms but rather in light of the nature and purpose of the contested measure.} Equality is a polysemous legal principle and shifts in the Court's choice of normative baseline in comparisons are difficult to predict in the absence of an explicit reference framework. 

This is further complicated as social advancements cause societal norms to shift. This fuels the difficulty of defining what a protected characteristic is. Protected characteristics fulfil a double function. On the one hand, they resemble and signal identity categories. On the other hand, in discrimination law, they serve as proxies for historical privileges and disadvantages. In other words, within society, particular groups of people have been disadvantaged in social arrangements and to account for historical injustice, these groups are afforded legal protection. As the boundaries of privileged and non-privileged might shift across contexts, different groups can be considered socially salient in different scenarios. 

\section{The Law is Not a Decision Tree}
\label{sec:discussion}
While algorithmic bias is not yet explicitly regulated, such regulation is likely to be adopted within a few years.\footnote{A proposal for an EU AI Act is currently under discussion at EU level.} This in turn raises the question of bias management and responsibility for unlawful algorithmic bias and unfairness. What is required of AI system providers to avoid or mitigate bias and when can AI system providers be said to have fulfilled this requirement? What limitations of current non-discrimination law should new regulations address? In this section, we discuss the implications of our findings.

When thinking about the law, many people envision some kind of tree structure, comprising of main rules and exceptions to those rules. While to some extent statutory law can be encoded as a decision tree, the analogy does not hold up to scrutiny. Instead, the law is dynamic, open-textured, and based on holistic reasoning. With regard to non-discrimination law in particular, (implicit) normative reasoning plays a fundamental role and the court rarely relies on statistical pointers. Further adding to this complexity, non-discrimination law is a polysemous legal instrument \cite{xenidis2021polysemy}. It fulfils a host of different social functions, ranging from the recognition of historical injustices and disadvantaged social groups, the (re)distribution of valuable goods and opportunities, the protection of dignity and autonomy, the accommodation of different lifestyles, and the facilitation of access to, and participation in, central social institutions such as the market, labour, education, healthcare, etc.\footnote{Many different scholars have reflected on this question: \citep{fraser1999social,honneth1996struggle,fraser2003redistribution,fredman2016substantive,mackinnon2016substantive}.} These various normative aims entail different conceptions of equality. While in a given context, formal equal treatment will suffice to fulfil the mandate of non-discrimination law, in others substantive or even transformative conceptions of equality will be required.

This suggests that, while many fairness metrics have taken inspiration from non-discrimination law, legal compliance cannot translate into a single threshold or fairness metric. Rather, fulfilling the requirements of non-discrimination law demands reflecting explicitly on the normative goal of legal and technical fairness interventions. Not doing so would render the notions of equality and fairness tautological \citep{westen1982empty}. In other words: focus should be shifted from questions such as "what should be the value of my fairness metric" to the more difficult yet crucial question of \textit{why} a particular distribution of burdens and benefits is right in a given context, and ultimately, \textit{who} should bear the costs of inequality. 

To assist practitioners in these endeavours, future work is necessary to uncover the moral implications of design choices in the machine learning development process. While discourse regarding the suitability of fairness metrics has received much attention in the legal community \citep[e.g.][]{hellman2020measuring,wachter2020bias,wachter2021fairness}, 
lawyers often have an idealised view of what fair-ml techniques can achieve~\citep{balayn2021beyond} and legal scholars have only recently begun to address the question of lawfulness of particular fair-ml strategies~\citep[e.g.][]{hellman2020measuring,kim2022race}. Understanding when particular interventions are appropriate is especially important considering the difficulties applicants face in providing \textit{prima facie} evidence in the context of opaque algorithmic systems.

\section{Conclusion}
\label{sec:conclusion}
In this paper, we set out to build a bridge between two separate disciplines: computer science and law. We analysed three seminal cases of algorithmic unfairness through the lens of EU non-discrimination law and showed that while the law offers protection against some types of algorithmic bias and unfairness, not all types of algorithmic unfairness neatly fall within the law's concepts and analytical frameworks. Subsequently, we explored the role fairness metrics can play in establishing legal compliance. In particular, we uncovered the normative assumptions of fairness metrics and the fair-ml algorithms that optimise for them and compared these to the legal reasoning of the Court of Justice of the EU. This analysis leads us to suggest that future research should inquire into what gets `lost in translation' when discrimination law as it is operationalised in judicial interpretation is expressed in terms of algorithmic (un)fairness and \textit{vice versa}. This would also entail a broadening of the scope of inquiry:
in order to meaningfully answer the question that non-discrimination law poses, we must move beyond merely asking \textit{what} should be equal and, instead, ask ourselves \textit{why} a particular distribution of burdens and benefits is right. 

\begin{acks}
This project has received funding from the European Union’s Horizon 2020 research and innovation programme under the Marie Skłodowska-Curie grant agreement No 898937.
\end{acks}

\bibliographystyle{ACM-Reference-Format}
\bibliography{references}

\newpage
\appendix

\end{document}